\documentclass[aps,prd,email,preprint,showpacs,showkeys,preprintnumbers,amsmath,amssymb,nofootinbib]{revtex4-1}

\usepackage{color}
\usepackage{latexsym}
\usepackage{amsmath}
\usepackage{amssymb}
\usepackage{eufrak}
\usepackage{euscript}
\usepackage{pstricks}
\usepackage{graphics}
\usepackage{graphicx}
\usepackage{picture}
\def\[{\left\lbrack}
\def\]{\right\rbrack}

\def\({\left(}
\def\){\right)}

\newcommand{\bbe}{\begin{equation}}
\newcommand{\eee}{\end{equation}}
\newcommand{\eaa}{\end{eqnarray}}
\newcommand{\baa}{\begin{eqnarray}}

\begin{document}

\pagestyle{myheadings}
\markright{Nonextensive statistics and entropic gravity applications........}

\title{\Large{Nonextensive statistics, entropic gravity and \\ gravitational force in a non-integer dimensional space}}

\author{Everton M. C. Abreu$^{a,b}$}
\email{evertonabreu@ufrrj.br}
\author{Jorge Ananias Neto$^{a}$}
\email{jorge@fisica.ufjf.br}
\author{Cresus F. L. Godinho$^b$}
\email{crgodinho@ufrrj.br}

\affiliation{$^a$Grupo de F\' isica Te\'orica e Matem\'atica F\' isica, Departamento de F\'{\i}sica, \\
Universidade Federal Rural do Rio de Janeiro\\
BR 465-07, 23890-971, Serop\'edica, RJ, Brazil \\\\ 
$^b$Departamento de F\'{\i}sica, ICE, Universidade Federal de Juiz de Fora,\\
36036-330, Juiz de Fora, MG, Brazil\\\\
\today\\}


\begin{abstract}
\noindent  Based on the connection between Tsallis nonextensive statistics and fractional dimensional space, in this work we have introduced, with the aid of Verlinde's formalism, the Newton constant in a fractal space as a function of the nonextensive constant.  With this result we have constructed a curve that shows the direct relation between Tsallis nonextensive parameter and the dimension of this fractal space. We have demonstrated precisely that there are ambiguities between the results due to Verlinde's approach and the ones due to fractional calculus formalism.  We have shown precisely that these ambiguities appear only for spaces with dimensions different from three.  Hence, we believe that this is a result in favor of our three dimensional world.
\end{abstract}


\maketitle

\pagestyle{myheadings}
\markright{Nonextensive statistics and entropic gravity applications........}






\section{Introduction}
\renewcommand{\theequation}{1.\arabic{equation}}
\setcounter{equation}{0}

The idea that gravity can be originated from thermodynamics first principles has begun with the discovery that black hole physics is connected to the laws of  thermodynamics \cite{1,2}.  These concepts were strongly boosted after Jacobson's work \cite{3}, where the Einstein equations were obtained from general thermodynamical approaches.  

In a recent work, Padmanabhan \cite{4} obtained an interpretation of gravity as an equipartition law.  In Verlinde's thermogravitational formalism \cite{verlinde}, the temperature and the acceleration are connected via Unruh effect \cite{unruh}.  At the same time, he combined the holographic principle with an equipartition law, where the number of bits is proportional to the area of the holographic surface.  Bits were used to define the microscopic degrees of freedom.  With these ingredients, the entropic force combined with the holographic principle and the equipartition law, the Newton's law of gravitation was obtained.

One possible interpretation of Verlinde's result is that gravity is not an underlying concept, but an emergent one.  It originates from the statistical behavior of the holographic screen microscopic degrees of freedom.  Following these ideas, the current literature has grown in an accelerated production from Coulomb force \cite{hs,wang} and symmetry considerations of entropic force \cite{8} to cosmology \cite{9} and loop quantum gravity \cite{10}, to mention some of them.

Considering the clear statistical origin of these entropic ideas, two of us, recently, used an extension of the standard Boltzman-Gibbs (BG) theory, which is well known as Tsallis' nonextensive statistics \cite{11} to analyze cosmological issues and to determine new bounds for the Tsallis parameter through noncommutative aspects \cite{12}.  Tsallis expression has a constant $q$ parameter which is also known as Tsallis $q$-parameter.
The Tsallis $q$-parameter measures the degree of nonextensivity.  The nonextensive formalism has been successfully used in a huge number of human knowledge areas besides physics.  It is important to realize, in advance, that when $q\rightarrow 1$ the usual BG theory is recovered.

Concerning classical systems, there are various problems that are not trivial which solution demands non-standard ways of calculation.  One of these systems is the one which encompasses nonconservative systems.   There are non-standard ways of calculation to attack this problem, and one of them is the fractional calculus (FC) which is one of the generalizations of the classical calculus.  It promotes a  redefinition of the mathematical tools and it is very useful to deal with anomalous and frictional systems \cite{13}, gravitational \cite{gravfrac,turcos}, constrained systems \cite{ag} and etc..   Besides, the investigation of non-integer dimensional spaces can bring us a representation of an effective physical analysis of confinement in low-dimensional systems \cite{he}.  To clarify, the fractional dimensional model permits us to analyze anisotropic excitations dynamics by solving the Schr\"odinger equation in a non-integer dimensional space, where the excitations are embedded in an isotropic environment.

However, it is a very interesting fact that Tsallis $q$ nonextensive statistics can be related to fractal extension for dynamics of complex systems \cite{frac1000} in such a way that both can be considered an unified new theoretical structure that can construct a proper base for the modern analysis of space plasma.  Besides, the $q$-formalism is connected to the fundamental fractal dynamics of the non-equilibrium states.

The fact that the nonextensive statistics can be connected to coarse grained (fractal) spaces \cite{12,aa} motivated us to use the fractal calculus to establish the behavior of the $q$-parameter in a non-integer dimensional space.  In this work we calculate the value of the non-extensive parameter through FC and, in a fractal space, we calculate the main values of the entropic gravity formalism as functions of the fractional dimension.

The organization of this paper follows a structure where the sections II and III are dedicated to introductory explanations
of Verlinde's and Tsallis' formalisms, respectively.  In section IV, after a very brief description of the important ingredients of the FC relevant results that will be used here, we will perform the connection between the  
fractal $\alpha-$dimensional space and  the value of the nonextensive parameter.  After that we will analyze the behavior of the gravitational constant in an $\alpha-$fractional space.  In section V we will discuss some ambiguities that appear in the fractional gravitational constant obtained by two different approaches that are the FC \cite{turcos,13,gravfrac} and the generalized $d$-dimensional Verlinde formalism \cite{RMJM}.   In section VI we will discuss the results obtained and the respective conclusions.

\section{A Brief Review of Verlinde's Formalism}
\label{vf}
\renewcommand{\theequation}{2.\arabic{equation}}
\setcounter{equation}{0}

One of the reasons that the study of entropy has been an interesting task through the recent years is the fact that it can be considered as a measure of information loss concerning the microscopic degrees of freedom of a physical system when depicting it in terms of macroscopic variables.  Appearing in different scenarios, entropy can be deemed as a consequence of the gravitational framework \cite{nicolini}.

The formalism proposed by E. Verlinde \cite{verlinde} derives the gravitational acceleration  by using, basically, the holographic principle and the equipartition law of energy. His ideas relied on the fact that gravitation can be considered to be universal and independent of the details of the spacetime microstructure \cite{nicolini}.  Besides, he brought new concepts about holography since the holographic principle must unify matter, gravity and quantum mechanics \cite{rev}.

The model considers a spherical surface as the holographic screen, with a particle of mass $M$ positioned in its center. A holographic screen can be imagined as a storage device for information. The number of bits (the term bit means the smallest unit of information in the holographic screen) is assumed to be proportional to the area $A$ of the holographic screen
\begin{eqnarray}
\label{bits}
N = \frac{A }{l_P^2},
\end{eqnarray}
where $ A = 4 \pi r^2 $ and $l_P = \sqrt{\frac{G\hbar}{c^3}}$ is the Planck length and $l_P^2$ is the Planck area.   In Verlinde's formalism we assume that the total energy of the bits on the screen is given by the equipartition law of energy

\begin{eqnarray}
\label{eq}
E = \frac{1}{2}\,N k_B T.
\end{eqnarray}
It is important to mention here that the usual equipartition theorem, Eq.(\ref{eq}), is derived from the usual BG thermostatistics. We will see that in a nonextensive thermostatistics scenario, the equipartition law of energy will be modified in a sense that a nonextensive parameter $q$ will be introduced in its expression.
Considering that the energy of the particle inside the holographic screen is equally divided by all bits then we can write the equation

\begin{eqnarray}
\label{meq}
M c^2 = \frac{1}{2}\,N k_B T.
\end{eqnarray}
Using Eq. (\ref{bits}) and the Unruh temperature formula \cite{unruh}

\begin{eqnarray}
\label{un}
k_B T = \frac{1}{2\pi}\, \frac{\hbar a}{c},
\end{eqnarray}
we are  in a position to derive the  (absolute) gravitational acceleration formula

\begin{eqnarray}
\label{acc}
a &=&  \frac{l_P^2 c^3}{\hbar} \, \frac{ M}{r^2}\nonumber\\ 
&=& G \, \frac{ M}{r^2}.
\end{eqnarray}
We can observe from Eq. (\ref{acc}) that the Newton constant $G$ is just written in terms of the fundamental constants, $G=l_P^2 c^3/\,\hbar$.

\section{The Tsallis Thermostatistics}
\label{TTS}
\renewcommand{\theequation}{3.\arabic{equation}}
\setcounter{equation}{0}

An important formulation of the nonextensive (NE) Boltzmann-Gibbs thermostatistics has been proposed by 
Tsallis  \cite{11} in which the entropy is given by the equation

\begin{eqnarray}
\label{nes}
S_q =  k_B \, \frac{1 - \sum_{i=1}^W p_i^q}{q-1}\;\;\;\;\;\; (\sum_{i=1}^W p_i = 1),
\end{eqnarray}
where $p_i$ is the probability of the system to be in a microstate, $W$ is the total number of configurations and $q$, 
as it was explained before, is the Tsallis parameter, or nonextensive parameter, or $q$-parameter.  It is a real parameter quantifying the degree of nonextensivity. 
The definition of entropy (\ref{nes}) has as motivation to study multifractals systems and it also possesses the usual properties of positivity, equiprobability, concavity and irreversibility.
It is important to note that Tsallis' formalism contains the BG statistics as a particular case in the limit $ q \rightarrow 1$ where the usual additivity of entropy is recovered. Plastino and Lima  \cite{PL}, by using a generalized velocity distribution for free particles  \cite{SPL} given by

\begin{eqnarray}
f_0(v) = B_q \[ 1-(1-q) \frac{m v^2}{2 k_B T} \]^{1/1-q},
\end{eqnarray}
where $B_q$ is a $q$-dependent normalization constant, $m$ and $v$ is a mass and velocity of the particle, respectively, they have derived a nonextensive equipartition law of energy whose expression can be written as,

\begin{eqnarray}
\label{ge}
E = \frac{1}{5 - 3 q} N k_B T,
\end{eqnarray}
where the range of $q$ is $ 0 \le q < 5/3 $.  For $ q=5/3$ (critical value) the expression of the equipartition law of energy, Eq. (\ref{ge}), diverges. So, it is obvious that $q=5/3$ is a forbidden value for $q$.  This fact will be reobtained numerically together with the fractional analysis in the next section.   Besides, it is easy to observe that for $ q = 1$,  the classical equipartition theorem for each microscopic degrees of freedom is recovered. It is important to mention that the virial theorem is not modified in this nonextensive  thermostatistics formalism  \cite{MPP}.

As an application of the nonextensive equipartition theorem in Verlinde's formalism we can
substitute the equipartition law by the nonextensive equipartition formula, Eq. (\ref{ge}), into Eq. (\ref{meq}), and applying the same steps described in section \ref{vf}, we can obtain a modified acceleration formula given by

\begin{eqnarray}
\label{accm}
a = G_{NE} \, \frac{ M}{r^2},
\end{eqnarray}
where $G_{NE}$ is an effective gravitational constant which is written as

\bbe
\label{S}
G_{NE}=\,\frac{5-3q}{2}\,G\,\,,
\eee
where we have reproduced the result in \cite{jorge}.
From the result (\ref{S}) we can observe that the effective gravitational constant depends on the nonextensive parameter $q$. For example, $q=1$ we have $ G_{NE}=G$ and for $q=5/3$ we have a curious and hypothetical result that is $G_{NE}=0$, which is, of course, unphysical as seen in Eq. 
(\ref{ge}) which diverges for $q=5/3$.   The fact that the value for $q$ can be also connected with fractal dimensions \cite{tsa} will reinforce $q = 5/3$ as a new forbidden value for $q$ and consequently establishes new bounds for future nonextensivity analysis.  As we said before, this result will be reobtained through a different approach in the next section.  It will be interesting to see how the fractional approach is connected to nonextensivity.  The noninteger (fuzzy) space is directly entangled with noncommutativity.  Verlinde's formalism states that gravity is emergent from entropy.  Since it can be shown that  gravity is emergent from noncommutativity, one can ask if gravity can be also emergent from nonextensivity.


\section{Nonextensive statistics in non-integer spaces}
\label{nonex}
\renewcommand{\theequation}{4.\arabic{equation}}
\setcounter{equation}{0}
\label{Res}

As we said in the Introduction section, the FC is extensively used in several areas that consider nonconservative systems.  However, we are more interested in its geometrical property since it can represent spaces with noninteger dimensions.

In the context of Gauss law in $\alpha-$dimensional fractional space \cite{turcos}, the total flux on the sphere is given by

\begin{eqnarray}
\label{TF}
\oint \vec{g}\cdot d\vec{A}=\oint g_{radial} dA=\frac{2\pi^\frac {\alpha -1}{2}}{\Gamma({\frac {\alpha -1}{2}})} \int_0^\pi \[ -\frac{\partial\varphi^\alpha}{\partial r} \] r^{\alpha-1} d\theta \sin^{\alpha-2}\theta\nonumber\\ \nonumber \\
=- \frac{m G_\alpha \alpha (\alpha-2) \pi^\frac{\alpha}{2}}{\Gamma(\frac{\alpha}{2}+1)}=-4\pi G m.
\end{eqnarray}
From (\ref{TF}) we can identify the Newton constant  in an $\alpha$-fractional space as

\bbe
\label{A}
G_{\alpha}\,=\,\frac{2\Gamma (\frac \alpha2)}{\pi^{\frac \alpha2 -1}(\alpha -2)}\,G,
\eee
where $G$ is the well known Newton's constant in Euclidean space.  
Hence, for $\alpha =  3$ in
(\ref{A}) we have $G_3=G$.  At the same time, $\alpha=3$ is the minimum integer value of $\alpha$ in order to keep physical (positive) the values of Newton's constant.  Conveniently, (\ref{A}) can be written as

\bbe
\label{B}
G_{\alpha}\,=\,\left[ \frac{2\,\Gamma(\frac{\alpha}{2})}{\pi^{\frac{\alpha}{2} -1}(\alpha -2)} \right]\,G_3 \qquad ,\qquad\qquad \alpha > 3
\eee
which indicates that all the other constants can be obtained from $G_3$ and consequently $G_{\alpha}$ can be understood as the underlying gravitational constant in this fuzzy (fractal) space with $D=\alpha$.  In the near future, we will relate this fractal gravitational constant with nonextensive values.

In figure 1 we have designed the curve that represents Eq. (\ref{B}). This curve has a minimum point. But it is not zero. This result could be naively interesting because the equation (4.3) has a $\pi^{\frac{\alpha}{2}-1}$ in the denominator. However, the gamma function behavior is the dominant one and that is the reason why this curve has this format. Besides, also due to the gamma function behavior, as the $\alpha$-dimension grow for small values of the interval, $G_\alpha/G_3$ is small too. And after an almost constant value (a plateau-like), $G_\alpha/G_3$ grows again for high dimensional fractal spaces.

\begin{figure}[!h]
\includegraphics[width=2.41in,height=2.32in]{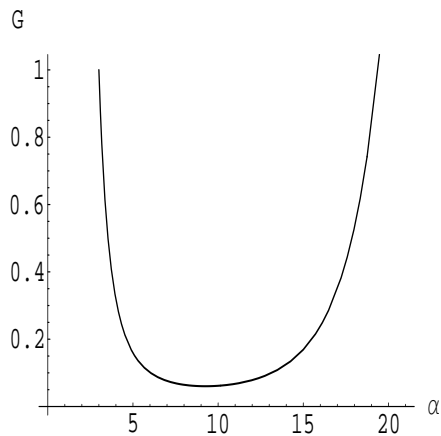}
\caption{Gravitational constant as function of $\alpha$ and $G=G_{\alpha}/G_3$}
\end{figure}

On the other hand, from \cite{jorge} we have that the nonextensive universal constant is given by

\bbe
\label{C}
G_{NE}\,=\,G\,\frac{5-3q}{2}\,=\,\left[ 1\,+\,\frac 32 (1-q) \right]\,G.
\eee

The fact that $q$ can be associated with fractality \cite{frac1000} motivates us to assume that there is a direct connection between $\alpha$ and $q$.   The fractal formalism used here provide us with the relation in Eq. (\ref{A}).   So, we will assume here that the $\alpha-q$ connection will be given by the Eq. (\ref{A})-(\ref{C}) connection such as

\bbe
\label{D}
\frac 32 (1-q)\,=\,\frac{2\,\Gamma(\frac \alpha 2)}{\pi^{\frac \alpha 2 -1}(\alpha -2)} -1,
\eee
or, in other words, we have that the $\alpha -q$ connection given by $G_{NE}=G_{\alpha}$.   From (\ref{D}) we can write that

\bbe
\label{E}
1\,-\,q\,=\,\frac 23\, \left[\frac{2\,\Gamma(\frac \alpha 2)}{\pi^{\frac \alpha 2 -1}(\alpha -2)} -1 \right],
\eee
which allows us to estimate a value for the $q$-parameter as a function of $\alpha$.  Other estimates for the $q$-parameter can be found in \cite{12} (and references therein) using other approaches such as noncommutativity.  Here we will use fractal calculus point of view to obtain such value.

To estimate precisely the value of $q$ from Eq. (\ref{E}) we have that

\bbe
\label{F}
q\,=\,\frac 53 \,-\,\frac 43\,\frac{\Gamma(\frac{\alpha}{2})}{\pi^{\frac{\alpha}{2}-1}(\alpha-2)},
\eee
which helps us construct the graphic in figure 2 where we can see clearly that the value $q=5/3$ never appears.   This is a consequence of Eq. (\ref{F}) where the second term in the right side never goes to zero.  This figure shows us also that we have a finite and relatively low number of dimensions since the denominator of the second term will never be too big in order to zero the fraction.  

One very interesting result is the value of $\alpha$ in order to recover BG statistics, i.e., when $q=1$.  In Eq. (\ref{E}) we can see that the l.r.s. of the equation has to be zero in order to obtain $q=1$.  It is easy to calculate that $\alpha=3$ for $q=1$.   Moreover, it is easy to see that there is another $\alpha$ for which we have $q=1$, i.e., when we recover BG statistics.  Since numerical computation is out of the scope of this paper, we can estimate $\alpha$ only graphically as being a non-integer value around 20.  Hence, it shows that $\alpha$ is non-integer in this fractal space $(D=\alpha)$ for the BG theory.
Another graphical result is the one for which $q=0$.  This result is important as being the minimal physical allowed valued for $q$, as we said above.  Note that in figure 2 we have omitted the coordinate $\alpha=2$ which causes a divergence.

However, in figure 3 we see clearly that $\alpha =2$ is  the divergent point.  Hence, we have the region between $\alpha =0$ and $\alpha =1$ which obeys the gamma functions behavior.  Then we have the infinities in $\alpha =2$ and after that, the same behavior shown in figure 2.

\begin{figure}[!h]
\includegraphics[width=3.41in,height=3.32in]{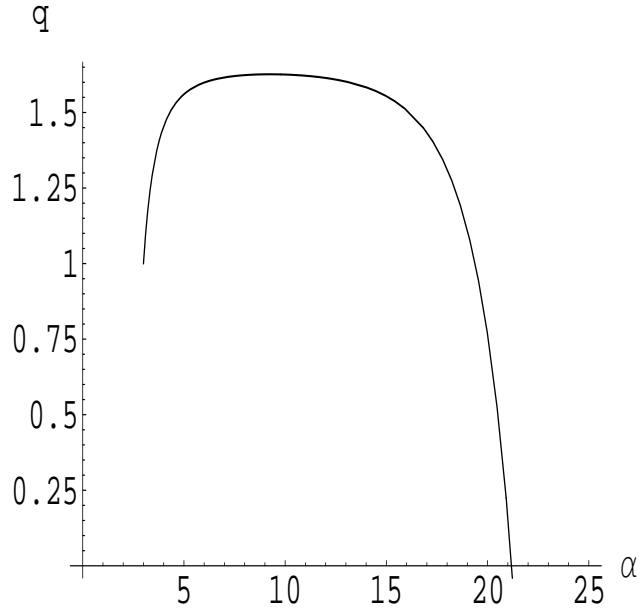}
\caption{$q$ behavior as function of $\alpha$}
\end{figure}

\begin{figure}[!h]
\includegraphics[width=2.41in,height=2.32in]{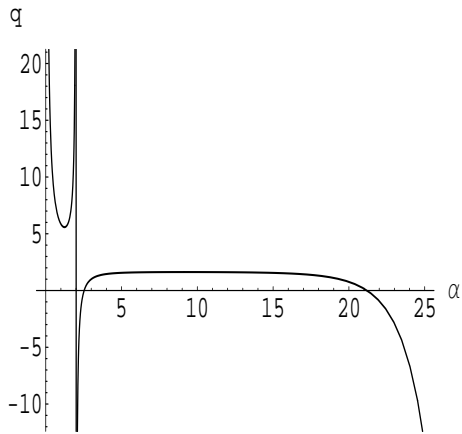}
\caption{Divergent behavior of $q$ for $\alpha=2$}
\end{figure}

To calculate $\alpha$ in order to obtain $q=0$ we have to use Eq. (\ref{F}) and solve the equation

\bbe
\label{134}
\pi^{\frac{\alpha}{2}-1}\,=\,0.8 \,\Gamma\Big(\frac{\alpha}{2} \Big)\,\frac{1}{\alpha-2},
\eee
which again has to be numerically calculated.  Graphically speaking for $q=0$, we have that $\alpha$ is a fraction after $21$ and so it is clear that $\alpha$ is noninteger.

Using Eq.(\ref{B}), we can also obtain the fractional force which is written as

\bbe
\label{g12}
F_{\alpha}\,=\,G_{\alpha}\frac{mM}{r^{\alpha-1}},
\eee

\noindent but from Eq. (\ref{B}) we can write (\ref{g12}) as

\bbe
\label{h12}
F_{\alpha}\,=\,G_3\,m\,M\,\frac{2\Gamma(\frac{\alpha}{2})}{\pi^{\frac{\alpha}{2}-1}(\alpha-2)\,r^{\alpha-1}}\,\,,\qquad \qquad \alpha > 3\,\,,
\eee

\noindent where it is easy to see that for $\alpha=3$ we have Newton's gravity in Euclidean space, namely

\bbe
\label{i12}
F_3\,=\,G_3\frac{m\,M}{r^2}\,\,,
\eee

\noindent which, substituting in Eq. (\ref{h12}) we have that

\bbe
\label{j12}
F_{\alpha}\,=\,\frac{2\Gamma(\frac{\alpha}{2})}{\pi^{\frac{\alpha}{2}-1}(\alpha-2)\,r^{\alpha-1}}\,F_3\,\,, \qquad \alpha > 3\,\,\,.
\eee

Hence, from Eq. (\ref{j12}) we can analyze the behavior of 
$F_{\alpha}$ when $r$ is kept constant ($r=1.2$, for instance) or when both $\alpha$ and $r$ are variables.  The first case numerical results are shown in figure 4 where the gravitational force has a minimum (a plateau-like) after a decay as we are traveling through fractal spaces of low dimensions.  After that, the force grows together with the fractional dimension following the behavior of the gamma function.  The next step is to vary $r$ and $\alpha$, which generates, obviously, a three dimensional graphic.  The result is given in figure 5 where we can see the valleys defined individually in figure 4 for each $r$-sector.  We can see also that as we travel to high dimensional fractal spaces the gravitational force grows rapidly.

\begin{figure}[!h]
\includegraphics[width=2.41in,height=2.32in]{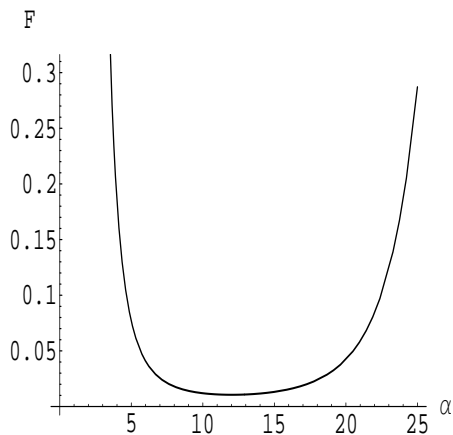}
\caption{Gravitational force behavior as function of $\alpha$ with constant $r=1.2$ and $F=F_{\alpha}/F_3$}
\end{figure}

\begin{figure}[!h]
\includegraphics[width=2.41in,height=2.32in]{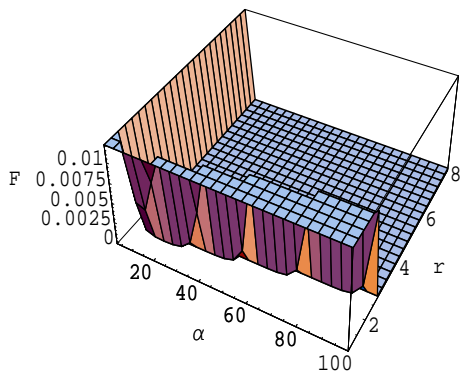}
\caption{Gravitational force behavior as function of $\alpha$ and $r$ and $F=F_{\alpha}/F_3$}
\end{figure}

We can see clearly in figure 5 the divergent sector that appears in figure 3 when $\alpha=2$.  It is shown in the first valley.  Notice that this first valley has a strange shape very different from the next ones.  The negative part of the divergence is covered by the blue plateau.

\section{Fractional Ambiguity in the Gravitational Constant}
\label{amb}
\renewcommand{\theequation}{5.\arabic{equation}}
\setcounter{equation}{0}

In this section we want to talk about different results for the gravitational constant that appear from different approaches described in this paper.  However, we will see that this ambiguity occurs only when $\alpha \neq 3$ which is a good result in favor of a $D=3$ space.  Speaking in a different way, we can say that this section could be considered as a demonstration of the number of spatial dimensions of the world we live.

Let us begin with the well known result that the fractional area is given by \cite{13}
\bbe
\label{a12}
\int dA\,=\,\int\,r^{\alpha-1}\,(\sin\theta)^{\alpha-2}\,d\theta d\phi\,\,,
\eee

\noindent where for $\alpha=3$ we obtain the Euclidean area.  So, we can write for (\ref{a12}) that \cite{13}

\baa
\label{a13}
\int\,dA\,&=&\,r^{\alpha-1}\int d\phi \int (\sin\theta)^{\alpha-2} d\theta \nonumber \\
&=&r^{\alpha-1}\,2\,\pi^{\frac{\alpha-1}{2}}\frac{1}{\Gamma(\frac{\alpha-1}{2})}\,\sqrt{\pi}\,\Gamma\Big(\frac{\alpha-1}{2}\Big)\frac{1}{\Gamma(\frac \alpha 2 )}\nonumber \\
&=&r^{\alpha-1}\frac{2 \pi^{\frac \alpha 2}}{\Gamma(\frac \alpha 2)},
\eaa

\noindent and the area of a fractional sphere will be

\bbe
\label{a14}
A_{\alpha}\,=\,\frac{2\pi^{\frac \alpha2}}{\Gamma(\frac \alpha2)}r^{\alpha-1},
\eee

\noindent for $\alpha=3$ it is easy to see that $A=4\pi r^2$.

From Verlinde's formalism described in section 2 we have that the fractional number of bits will be

\bbe
\label{a15}
N_{\alpha}\,=\,\frac{A_{\alpha}}{l^{\alpha-1}_P}\,=\,\frac{2\pi^{\frac{\alpha}{2}}}{\Gamma(\frac{\alpha}{2})}\,\frac{r^{\alpha-1}}{l^{\alpha-1}_P},
\eee

\noindent but we also have that, from Verlinde's approach

\baa
\label{b12}
Mc^2&=&\frac 12 \,N_{\alpha} \kappa T \,=\,\frac 12\,N_{\alpha} \frac{1}{2\pi} \frac{\hbar a}{c} \nonumber \\
\Longrightarrow a&=&M\frac{c^3 l^{\alpha-1}_P}{\hbar}\frac{2\Gamma(\frac{\alpha}{2})}{\pi^{\frac{\alpha}{2} -1}r^{\alpha-1}} \nonumber \\
\Longrightarrow a&=& G_{\alpha} \frac{M}{r^{\alpha-1}},
\eaa

\noindent where 

\bbe
\label{b13}
G_{\alpha}= 2\pi^{1-\frac{\alpha}{2}}\Gamma(\frac{\alpha}{2})\frac{c^3 l^{\alpha-1}_P}{\hbar},
\eee

\noindent which is the same result derived in \cite{RMJM}. 

In order to compare the gravitational constants obtained in sections (\ref{nonex}) and (\ref{amb}), we first rewrite the gravitational constant obtained by Verlinde's formalism, Eq.(\ref{b13}), as

\begin{eqnarray}
G_{\alpha}^{VF}=\frac{2\Gamma(\frac{\alpha}{2})\frac{c^3 l^{2}_P}{\hbar}}{\pi^{\frac{\alpha}{2}-1}}
\;l^{\alpha-3}_P,\nonumber 
\end{eqnarray}
and consequently we have that

\begin{eqnarray}
\label{b132}
G_{\alpha}^{VF}=\[\frac{2\Gamma(\frac{\alpha}{2})}{\pi^{\frac{\alpha}{2}-1}}\]\, \,l^{\alpha-3}_P\, G_3,
\end{eqnarray}
where the superscript VF means Verlinde's formalism and $G_3\equiv\frac{c^3 l^{2}_P}{\hbar}$. Rewriting the gravitational constant obtained by FC, Eq.(\ref{B}),  in a similar form of Eq. (\ref{b132}), we have
\bbe
\label{b133}
G_{\alpha}^{FC}=\[\frac{2\Gamma(\frac{\alpha}{2})}{\pi^{\frac{\alpha}{2}-1}\,(\alpha-2)}\]\, G_3.
\eee 
By looking at Eqs. (\ref{b132}) and (\ref{b133}), we can note that the first difference is the presence of the term $(\alpha-2)$ in the denominator of Eq. (\ref{b133}). The second, and maybe the most important difference, is the correction term, $\,l^{\alpha-3}_P$, in Eq. (\ref{b132}). This term, in principle, does not allow  to obtain a direct connection between the Tsallis parameter $q$ and the fractal dimension $\alpha$,  for $\alpha\neq 3$, in the context of Verlinde's formalism. This occurred because the nonextensive gravitational constant, Eq.(\ref{S}), has units of $G_3$ and the effective gravitational constant, Eq.(\ref{b132}), has units of $\,l^{\alpha-3}_P\, G_3$.  If we promote the equality $G_{NE}=G_{\alpha}^{VF}$ we can obtain  an algebraic relation between $q$ and $\alpha$ containing the term $l^{\alpha-3}_P$. This result is clearly inconsistent because both $q$ and $\alpha$ are dimensionless.
Unlike, the fractal or dimensional correction in Eq.(\ref{b133}), which was derived from FC, is dimensionless.  For $\alpha=3$, both equations bring the same value. This last result is also important because for $\alpha=3$ there is no ambiguity in different approaches for the 
$d$-dimensional gravitational constant. This fact may indicate that our space is indeed three-dimensional.

\section{Conclusions}

One of the greatest challenges of today's theoretical physics is the problem of unification of the fundamental forces that govern the Universe. The recent attempts to unify gravity and quantum mechanics are very popular. A well known candidate to promote the knowledge of the physics of the early Universe is the noncommutative geometry which, as a consequence of the position uncertainty principle, produce a kind of fuzzy space.  This fuzziness concept can be understood as a fractal property of spacetime.

Since theoretically it is well known that it can be possible that gravity can emerge from noncommutativity, i.e., from fuzziness, this result motivated us to try to fathom the consequences of this so-called non-integer dimensional space in some gravity ideas developed recently.
In this work we have chosen Verlinde's emergent gravity framework (which has statistical basis) to study these fractal concepts through the fractional calculus.

Besides, the statistical theory used here was the nonextensive one developed by Tsallis in the late eighties.  This formalism has shown several successes in many branches of our knowledge and we believe that it would fit into our needs here since the nonextensivity is also connected with fractal spaces.  In our opinion, the fact that Tsallis theory is not additive turn it into one of the best statistical choices that we can use when dealing with Verlinde's formalism, as it can be explicitly seen in \cite{aa}.

In this work we have analyzed the influence of the dimension of any fractal space in the gravitational constant.  We have constructed $D=2$  graphic in order to see the behavior of the gravitational constant as a function of the dimension $\alpha$.  In non-integer dimensional spaces the $\alpha$-gravitational constant behavior follows the gamma function behavior.

After that we have promoted a connection between these concepts in order to construct paths to compute the fundamental constants of gravity and Tsallis formalism, namely, the $q$-parameter which measure the degree of nonextensivity of the systems.  Concerning this quest we have constructed a relation between the dimension of this non-integer space and the $q$-parameter.  In other words, we could use the $\alpha$-dimension also to measure the nonextensivity.  A curve which shows the $q \times \alpha$ behavior was constructed.  We have shown that $q=5/3$ is a forbidden value for $q$ and consequently it can establish new bounds for $q$ values.  Besides, since superstrings and supergravity theories are showing us the importance of extra dimensions we also extended this analysis to our computations.   

Next we have analyzed the influence of the fractional dimension in the gravitational force.  We have constructed  $D=3$ graphic in order to see the behavior of the force as a function of the dimension $\alpha$ and the distance $r$ between two hypothetical masses.    In non-integer dimensional spaces the force behavior also follows the gamma function behavior.

Finally, we have detected that there is an ambiguity for $\alpha\neq 3$ between the $G$ results provided by the fractional calculus and the one calculated directed through the Verlinde approach.  This ambiguity indicates that our space has to be three-dimensional.  Hence, the superior dimensions that we worked in section IV have to suffer some kind of spatial reduction to $\alpha =3$.  This reduction process is out of the scope of this paper.  However, it can be, in some way, a demonstration that we live in a $D=3$ spatial dimensional world.

All these results can show us that all these concepts, i.e., non-integer space, Verlinde's entropic gravity formalism and the Tsallis thermostatistics concept are all entangled.  The objective here was to obtain new concepts involving these theories in order to show new points of views.

\section{Acknowledgments}

E. Abreu would like to thank CNPq (Conselho Nacional de Desenvolvimento Cient\' ifico e Tecnol\'ogico) and C. Godinho would like to thank FAPERJ (Funda\c{c}\~ao de Amparo \`a Pesquisa do Estado do Rio de Janeiro, for partial financial support. CNPq and FAPERJ are  Brazilian scientific support agencies.

\bigskip
\bigskip



\end{document}